# What Library Digitization Leaves Out: Predicting the Availability of Digital Surrogates of English Novels

Allen Riddell[†]        Troy J. Bassett[*]

First version: 22 August 2019
This version: 1 September 2020

### Abstract

Library digitization has made more than a hundred thousand 19th-century English-language books available to the public. Do the books which have been digitized reflect the population of published books? An affirmative answer would allow book and literary historians to use holdings of major digital libraries as proxies for the population of published works, sparing them the labor of collecting a representative sample. We address this question by taking advantage of exhaustive bibliographies of novels published for the first time in the British Isles in 1836 and 1838, identifying which of these novels have at least one digital surrogate in the Internet Archive, HathiTrust, Google Books, and the British Library. We find that digital surrogate availability is not random. Certain kinds of novels, notably novels written by men and novels published in multivolume format, have digital surrogates available at distinctly higher rates than other kinds of novels. As the processes leading to this outcome are unlikely to be isolated to the novel and the late 1830s, these findings suggest that similar patterns will likely be observed during adjacent decades and in other genres of publishing (e.g., non-fiction).

## 1 Introduction

Bulk library digitization has made hundreds of thousands of digital surrogates of 19th-century books available to researchers and the general public. Researchers are only beginning to explore possible uses of these surrogates. Book historians and bibliographers, for example, can inspect title pages and interiors of books of interest, eliminating the need to travel to distant libraries or even their own libraries.

---

[†]   [†]Indiana University Bloomington.
[*]   [*]Purdue University Fort Wayne.

Many of the more exciting potential uses of digital surrogates of 19th-century books held by major digital libraries (British Library, Google Books, HathiTrust, and Internet Archive) involve the analysis of book contents. Today it is possible to imagine reconstructing—with the aid of optical character recognition—a detailed view of, say, the range of themes, plots, and settings discussed in works of fiction published in the British Isles during the 19th century. But this project and similar projects depend on the population of books with digital surrogates resembling the larger population of published books. If these two populations do not resemble each other, then researchers learn only about the collections of libraries contributing digital surrogates. They learn nothing about the history of publishing in general.

In this study we look at the availability of digital surrogates of novels first published in the British Isles during the late 1830s. This is a period for which we have lists of every novel which was published. We ask whether the population of novels with digital surrogates resembles the population of published novels. (For example, is the share of novels written by women roughly the same in both populations?) We ask this question because the consequences of an affirmative answer are profound. If the two populations resemble each other, researchers can use novels which are already readily accessible to explore aggregate characteristics and trends in the history of publishing in the British Isles. That is, they need not worry that certain kinds of novels tend to be missing from the corpus of novels with digital surrogates.

## 2 Background

Starting around 2004 organizations including Google, Microsoft, and the Internet Archive digitized millions of volumes held by major university and state libraries.[1] Participating libraries included the University of California, Harvard University, Oxford University, and the British Library. Although books published during the 20th century are, in general, protected by intellectual monopolies of varying lengths, the digital surrogates of books published during the 19th century tend to be freely accessible online.

Questions about the "coverage" of digital libraries have dogged researchers wanting to make inferences about publishing activity during the 19th century based on available digital surrogates.[2] The hope that available digital surrogates might adequately reflect the population of books published during the 19th century is far from wishful thinking, especially if this hope is restricted to certain geographic areas, genres of publishing, or certain decades. Most titles published as books in the 19th century survive due to their physical durability and generous print runs (often in the thousands of copies). That most books survive is also due to legal deposit laws which mandate that publishers send copies of new books to at least one (state) library. In the case of the British Isles, two of the five legal deposit libraries (the British Library and Oxford University) participated in early library digitization efforts.[3] The participation of these legal deposit libraries in mass digitization projects makes plausible the proposition that, for this location and period, the population of available digital surrogates may approximate the population of editions published.

Digitization of general circulation collections of state and research libraries began in 2004 when the internet search and advertising company Google launched the Google Books Project.[4]

Significant early partnerships with Google Books included Oxford, Harvard, the University of Michigan, the New York Public Library, and Stanford University. Google Books would go on to scan more than thirty million volumes from various libraries.[5] Other libraries, working with the Open Content Alliance (OCA) and contracting with the non-profit Internet Archive for digitization services, began digitizing their collections in 2005. Early OCA participating libraries included the British Library, the University of Illinois, and the Boston Public Library.[6] Between 2005 and 2008 the OCA's digitization work received significant funding from the software company Microsoft as part of the Live Book Search service.[7] In 2013 the Internet Archive reported that it had digitized two million books.[8]

Library digitization continues in the present. Many large libraries, including several of those mentioned above, operate their own digitization programs. Libraries often make the digital surrogates they produce available on their own websites, often via links in the relevant records in their online catalogs. In North America, a copy of a newly created digital surrogate will likely be deposited at HathiTrust, a library consortium which accepts deposits of digitized volumes from member libraries.[9]

In the interest of concision, we will refer to the four major digital libraries for English-language books published during the 19th century (Internet Archive, HathiTrust, Google Books, and the British Library) as "the major digital libraries," omitting reference to their focus on English-language books and English-speaking regions of the world.[10]

Holdings of digital surrogates overlap. In many cases, a digital surrogate available from one major digital library is available from one or more additional digital libraries. HathiTrust, for example, makes available many—but not all—public domain digital surrogates which were created by Google Books.[11] As a consequence, these volumes are typically also available from Google Books. As the Internet Archive was commissioned by the British Library to digitize some of its collections, many of the copies digitized are available from both the Internet Archive and the British Library. Thanks largely to the efforts of Aaron Swartz, the Internet Archive includes copies of hundreds of thousands of digital surrogates available from Google Books.[12] Although overlap is considerable, there are always digital surrogates which are only available from one of the four major digital libraries. For this reason, checking each digital library separately is typically required to identify whether or not an English-language book has a publicly available digital surrogate. A researcher interested in locating a digital surrogate of an English-language 19th-century book will typically perform a search using the websites of HathiTrust, Internet Archive, and Google Books. If their initial search yields no results, they may use a general purpose search engine (e.g., Bing, Google, DuckDuckGo) and search specific library catalogs if they believe page images may be available at a library which is not connected with either Google Books, HathiTrust, or the Internet Archive. The British Library is an important example of such a library and many of its digital surrogates can be found only by using the library's online catalog.

Numerous academic articles have made use of digital surrogates from one or more of the major digital libraries. One important line of research in information and library science attempts to make use of the text of digital surrogates to advance bibliographic goals. (The "text" of a digital surrogate is the machine-readable text produced by optical character recognition (OCR).)

For example, Bamman, Carney, Gillick, Hennesy, and Sridhar (2017) use the text derived from digital surrogates held by HathiTrust to predict the date of first publication of a work.[13] As titles lacking (reliable) publication dates are relatively common during the 19th century, this research allows scholars to make inferences, supported by the words appearing in a text, about an undated volume's likely publication date.

Research more closely related to the material presented in this paper characterizes the coverage of one or more of the major digital libraries with respect to some (ideal) benchmark, often for a specific kind of book or document. Jones (2011) explores Google Books' coverage of 19th century books, using as a benchmark the five-volume *Catalogue of the Library of the Boston Athenæum, 1807–1871*.[14] Using a random sample ($N = 398$) from the Boston Athenæum catalog, Jones (2011) attempts to locate matching digital surrogates in Google Books, finding a match in 235 cases (59%).[15] Sare (2012) considers Google Books' and HathiTrust's coverage of US government documents published between 1943 and 1976.[16] Like Jones (2011), Sare (2012) uses a random sample ($N = 1540$) from the population of government documents. Sare (2012) finds that 436 documents (28%) were available in some form from HathiTrust and that 809 documents (53%) were available from Google Books.[17] Sare (2012)'s study is distinctive in that it makes use of an exhaustive list of items which clearly defines a population: every US government document sent to depository libraries between 1943 and 1979. Jones (2011), by contrast, uses a library catalog, leaving open the question of what books were published but never collected by this particular library. Although both approaches are extremely valuable, we use a method more similar to the one used by Sare (2012) because it avoids having to reason about missing items in a particular library collection.

## 2.1 Which Kinds of Books are Likely to have Digital Surrogates?

Anyone who has had occasion to search online for page images of books published during the 19th century knows that not every edition has a digital surrogate. Countless 19th-century editions survive in libraries which have not digitized their collections. Many libraries digitized some but not all 19th-century books in their collections. Many factors influence whether a given edition has a digital surrogate available in at least one of the major digital libraries. We have found it useful to group the obvious factors into two categories. Factors in the first category relate to *collecting practices*. These factors influence whether or not a library contributing to the major digital libraries is likely to have a copy of a given edition in its holdings. (A library possessing a copy of an edition is, of course, a prerequisite for the library's contributing a digital surrogate.) The existence of legal deposit requirements for publishers after 1710 give us reason to believe that collecting practices may not be particularly influential, at least for the legal deposit libraries contributing to the major digital libraries (Oxford and the British Library).[18] If publishers in the British Isles reliably sent copies of new books to Oxford and the British Library, then every 19th-century novel should, in principle, be found in the holdings of these two libraries. For these two libraries there may be no "collecting practices" worth mentioning. For North American libraries, by contrast, collecting practices may well influence which 19th-century English-language novels are in their holdings, especially those originally published in the British Isles.

Factors in the second category concern *digitization practices* at contributing libraries. Anecdotal experience with bulk digitization programs suggests that digitization programs did not, as a rule, selectively digitize certain books on library shelves. Shelves of books and floors of shelves were chosen for digitization, not individual books. But the shelves or floors of shelves selected for digitization may not have been chosen uniformly at random. Shelves in special collections tended to be skipped, for example. Books may also have been passed over for technical reasons. Digitization equipment used in bulk digitization programs is typically not able to process very large or very small items. Large-format books containing maps, for example, tended to be left on shelves.[19] Although the practice of passing over very large and very small items is not one we anticipate being relevant in the case of the 19th-century novel, it is an unambiguous case of digitization practices yielding an unrepresentative sample of library collections.

In this section we will discuss four features associated with a novel which may influence collecting practices or digitization practices: print run, (sub)genre, author gender, and format. Our initial thinking was that only (sub)genre influences both collecting and digitization practices. The remaining features seem likely to influence only collecting practices.

The print run of an edition seems very likely to influence whether or not a given edition has a digital surrogate available in one or more of the major digital libraries. Books with higher first-edition print runs were more likely to be collected than books with lower print runs. Such books tended to be more popular as high print runs tend to reflect a publishers' prospective judgment of demand. More popular books, all other things being equal, seem more likely to have been targets for library and private collection. Moreover, books with higher print runs would be more likely to survive in libraries because a greater number of copies implies (mechanically) that a greater opportunity for collection existed. Because the print run of a novel is typically unknown, print run seems likely to be only relevant to a consideration of collecting practices.[20]

The genre of a novel could also have influenced whether or not a given edition has a digital surrogate available. Genre may be a factor which influences collection practices and digitization practices. Certain genres may have been systematically targeted for collection during the 19th century. Libraries may have judged collecting works in certain genres (e.g., historical novels) more desirable. Genre could also influence digitization practices. If novels in certain genres were segregated and shelves containing them tended to be passed over (or preferentially selected), then a book's genre could influence whether or not a digital surrogate is available. For example, if works classified "juvenile fiction"—for whatever reason—were shelved in a different area than other works of prose fiction, we can imagine a scenario in which these shelves tended to be passed over. If libraries facing budget or time constraints were unable to scan all their holdings, we speculate that they might have selected one "type" of fiction rather than the other.

Format and author gender are two remaining factors which may have influenced collecting practices. (We have no reason to believe that they directly influenced digitization practices.) The format used for the first edition of a 19th-century novel was a rough indication of the work's prestige. Having a work published in a multivolume format (typically two or three volumes) was regarded as more desirable than having a work published in a single-volume format.[21] The distinction is vaguely analogous to the contemporary division between works whose first edition uses a hardcover format and those which use a paperback format. If libraries used format to guide their collecting practices, we would expect that format would predict a novel having a first edition digital surrogate. Author gender may also have influenced collecting practices. The prevalence and impact of bias against women novelists by text industry intermediaries (e.g., reviewers, publishers) during the 19th century has been the subject of sustained discussion.[22] Women novelists may have tended to write novels in specific genres (e.g., juvenile fiction). Novels in these genres may have been less likely to be targets for library collection. If those involved in library collection practices used gender as an indicator of or proxy for collection-worthiness then author gender may predict digital surrogate availability. In the discussion that follows, we will assess not only if format and author gender affect the availability of digital surrogates, but crucially if the intersection of format and author gender affect the availability as well. While individually format and author gender could each affect availability, together the factors could reinforce the effects of bias. If libraries are truly digitizing their collections at random, then there should be no discernable difference between the availabilities of digitized surrogates by format, author, or their combination.

This article focuses on two research questions:

1. Do the novels which have been digitized reflect the population of published novels? For example, is the share of novels written by women roughly the same in both populations?

2. If they do not, which kinds of novels are over- or under-represented?

## 3 Methodology

Our approach is easy to describe. We first gather a list of novels first published in the British Isles in 1836 and 1838. The list includes every first-edition novel published in 1838 and a simple random sample of first-edition novels published in 1836. For each novel in the population, we search the major digital libraries for a digital surrogate. If we find at least one digital surrogate, we record this fact.

The most important materials in this investigation are two exhaustive lists of novels. These lists contain records of novels published during 1836 and 1838. (1836 and 1838 are the most recent two years for which exhaustive bibliographies of novels exist.) These two lists enumerate the population of novels which could, in principle, have been digitized. Both these lists include author gender annotations and novel format information. Author gender typically matches the gender of the historical individual credited with authorship. (Appendix A describes the gender annotation procedure in detail.) T. Bassett (n.d.) provides an exhaustive list of the 94 novels published in 1838.[23] We gather information about digital surrogate availability for all these novels. Garside, Mandal, Ebbes, Koch, and Schöwerling (2004) provides an exhaustive list of the 90 novels published in 1836. This list is part of a larger (exhaustive) bibliography covering 1770 to 1836.[24] To economize on time, we sample uniformly at random (without replacement) from these novels ($n=32$).[25] We add this 1836 random sample to the list of 1838 novels primarily to address the remote possibility that text industry output during the year 1838 might have been exceptional in some unanticipated way.

In this study we use an inclusive definition of a novel. A novel is a work of prose fiction of at least 90 pages not addressed primarily to children under the age of 13. This is the definition used by T. Bassett (n.d.).[26] This definition is more inclusive than the one used by the standard bibliography covering 1770–1836 developed by Garside and collaborators. Garside and collaborators focused on works labeled as "novels" by contemporaries and therefore exclude prose fiction which they characterize as didactic-religious as well as prose fiction which they classify as juvenile fiction (but which likely addressed some adult readers). Our inclusive definition has the virtue of being easier to apply. It requires fewer judgments by domain experts. For example, the definition used by Garside and collaborators requires an expert to decide whether or not contemporaries referred to the book as a novel. Reaching a decision on this question might require consulting contemporary reviews and publisher advertisements. Switching between the two definitions for novels published during 1836 is made easy by the fact that Garside et al. (2004) lists works which fell short of their definition in appendices. To use the inclusive definition for 1836 works, we include works in these appendices which qualify under the more permissive definition.

Using the combined list of 126 novels published during the late 1830s, we search the major digital libraries for first edition digital surrogates. That is, we use the relevant web interfaces to search the holdings of the major digital libraries. We count a novel as having a digital surrogate if a digital surrogate exists for the first edition. If the novel is a multivolume novel all volumes must be available for the novel to count as having a digital surrogate. We count as first editions versions of the novel which were published in the same year as the first edition by the first edition publisher. For example, an "export edition" destined for Canada with a variant title page could count as a first edition digital surrogate if it was published in the same year as the first edition by the same publisher. We find that of the 126 novels, 106 (84%) have at least one digital surrogate available in the major digital libraries. Table 1 shows the counts of novels with digital surrogates by author gender and novel format.

|  | Has digital surrogate | Total |
| --- | --- | --- |
| Woman-author, single-volume | 19 (68%) | 28 |
| Woman-author, multivolume | 20 (91%) | 22 |
| Man-author, single-volume | 17 (85%) | 20 |
| Man-author, multivolume | 42 (95%) | 44 |
| Unknown-gender author, single-volume | 5 (56%) | 9 |
| Unknown-gender author, multivolume | 3 (100%) | 3 |
| All novels | 106 (84%) | 126 |

**Table 1:** Availability of digital surrogates of 1836 and 1838 novels by author gender and novel format.

## 3.1 Characterizing Uncertainty about Rates of Digital Surrogate Availability

To address our research questions we need to estimate credible intervals for the underlying rates at which novels have first edition digital surrogates during the late 1830s (1835–1839). In order to form reasonable beliefs about the rate of digital surrogate availability for years in the late 1830s we must write down a model which allows our observations from 1836 and 1838 to inform our beliefs about digital surrogate availability for the population of late 1830s novels. Such a model should also permit us to estimate the rate of availability for novels in different categories, where categories are defined by author gender and novel format. Were we to find rates which resembled each other, this would count as evidence in favor of the theory that the processes leading to a novel having a first edition digital surrogate are largely random.

To estimate the credible intervals for each category of novels, we use a hierarchical model with Beta priors and Binomial sampling distributions for each of the six categories of novels. In symbols,

$$\alpha \sim \text{Gamma}(2, 1)$$
$$\beta \sim \text{Gamma}(2, 1)$$
$$p_i \sim \text{Beta}(\alpha, \beta), i \in \{1, 2, \ldots, 6\}$$
$$y_i \sim \text{Binomial}(N_i, p_i) \, i \in \{1, 2, \ldots, 6\}$$

where $p_1, p_2, \ldots, p_6$ are the unobserved rates—which we estimate from the data—at which each category of novels has first edition digital surrogates, $N_1, N_2, \ldots, N_6$ are the number of novels in each category, and $y_1, y_2, \ldots, y_6$ are the number of novels we observe having available digital surrogates in each category. Our model treats the 1836 and 1838 observations as if they were randomly sampled from the population of late 1830s novels.[27] This assumption simplifies the model considerably. We think the decision to treat the observations as indistinguishable from observations in the population is reasonable. We know of no reason why the specific publication year of a late-1830s novel would influence collection or digitization processes. Moreover, we can perform rudimentary checks of this modeling assumption by verifying that digitization rates do not vary much by year. For example, the empirical proportion of 1836 woman-authored novels which have digital surrogates is very similar to the rate of women-authored novels in 1838 (75% and 79%) have digital surrogates.

The use of a model to characterize uncertainty is valuable even if it were restricted to the analysis of 1838 titles, a year for which our data includes every title published. In this case the uncertainty estimates can be understood as a description of the probability that a hitherto unacknowledged 1838 novel in a given category will be observed to have a digital surrogate. Novels previously unknown to bibliographers do occasionally surface. For example, a handful of novels first published between 1800 and 1829 were discovered after the publication of the landmark bibliography of Garside and Schöwerling (2000), a bibliography of the period believed to be exhaustive at its time of publication.[28] So a model characterizing the posterior probability of digital surrogate availability would be valuable even if we had observations of availability for all late-1830s novels.

We estimate the posterior probability distributions with Markov Chain Monte Carlo using the software Stan.[29] Code and data are published under the following DOI: 10.5281/zenodo.4010771.

# 4 Analysis

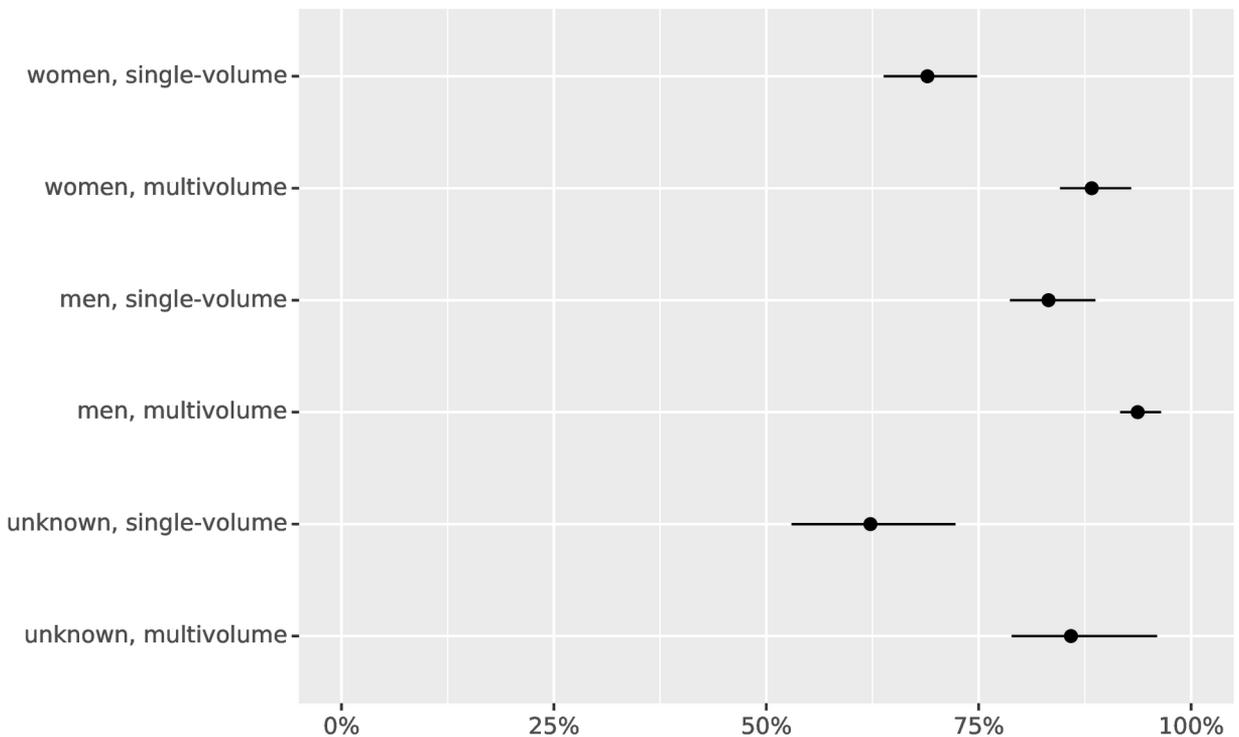

**Figure 1:** Estimated rates of digital surrogate availability ($p_1, p_2, \ldots, p_6$). Points indicate posterior means, bars show 50% credible intervals.

Digital surrogate availability varies by author gender and format. Posterior estimates of availability rates appear in Figure 1. These rates were estimated using the model and data described in the previous section. As described in the previous section, the model allows us to characterize uncertainty about the rates, something we cannot accomplish using the counts of available digital surrogates alone (Table 1). Although there is uncertainty about the rates, especially for the unknown-gender author novels, several differences are obvious. The mean estimate for the rate at which women-author, single-volume novels from the late 1830s have digital surrogates is 69%. The mean estimate for the rate for men-author, multivolume novels from the late 1830s is 94%. Taking account of uncertainty about the rates, it is is virtually certain (99% probability) that the latter rate is higher than the former. We are also virtually certain that the rate at which men-author, multivolume novels have digital surrogates is higher than the rate at which unknown-gender author, single-volume novels have surrogates.

Multivolume novels are more likely to have digital surrogates than single-volume novels. The rate at which multivolume novels by women have digital surrogates is very likely greater than the rate at which single-volume novels by women have digital surrogates. ("Very likely" means that its posterior probability of the inequality is greater than 90% but less than 99%.) The pattern holds true for multivolume and single-volume novels by men authors and by unknown-gender authors. The mean estimates of the rates of digitization for multivolume and single-volume novels by men are 0.94 and 0.83, respectively. The mean estimates of the format-specific rates for unknown-gender author novels are 0.86 and 0.62, respectively. Our relative lack of confidence in the difference between rates for unknown-gender authors—despite the large difference between the mean estimates—is due to the remaining uncertainty about the rates. This uncertainty is due to the small number of unknown-gender titles in the sample (10 out of 126). With so few observations there is only so much confidence we can gain about the rates of availability for these novels.

An alternative way of thinking about these findings is to consider what might happen if a researcher were to sample 200 first-edition late-1830s novels from the major digital libraries. Such 200-novel samples will tend to have about 74 novels by women, 37 of which would be single-volume novels. Had the samples been taken at random from the population of published novels, however, we would expect to see 79 novels by women, 44 of which would be single-volume novels. Hence sampling from the major digital libraries instead of the population will tend to undercount women-author, single-volume novels by about 17%. Single-volume novels by unknown-gender authors would be undercounted by 25%. Novels which are over-counted in this scenario are primarily multivolume novels by men. Samples from the major digital libraries will tend to over-represent this group by 13%.

## 5 Is Within-Library Digitization Random? The Case of the British Library

Our main finding is that different kinds of novels have digital surrogates at different rates. Format and author gender can be used to predict digital surrogate availability. This result makes using samples from the major digital libraries inadvisable, as these samples will not reflect the population of published novels. Yet the result raises a question: Why do we observe these patterns in the availability of novels? Are the differences due to library collecting practices? Or is there perhaps evidence that digitization practices are driving the differential availability of novels in different categories?

| | Has digital surrogate in British Library | Total |
|---|---|---|
| woman-author, single-volume, 1836 | 4 (44%) | 9 |
| woman-author, multivolume, 1836 | 0 (0%) | 10 |
| man-author, single-volume, 1836 | 6 (60%) | 10 |
| man-author, multivolume, 1836 | 0 (0%) | 25 |
| unknown-gender author, single- | 4 (67%) | 6 |
| volume, 1836 | | |
| unknown-gender author, multivolume, 1836 | 0 (0%) | 2 |
| woman-author, single-volume, 1838 | 13 (59%) | 22 |
| woman-author, multivolume, 1838 | 6 (38%) | 16 |
| man-author, single-volume, 1838 | 8 (47%) | 17 |
| man-author, multivolume, 1838 | 14 (42%) | 33 |
| unknown-gender author, single-volume, 1838 | 1 (25%) | 4 |
| unknown-gender author, multivolume, 1838 | 0 (0%) | 2 |
| All novels | 56 (36%) | 156 |

**Table 2:** Availability of digital surrogates from the British Library by year, author gender, and novel format. Note that no 1836 multivolume novels have been digitized. Totals differ from Table 1 because we check all 1836 novels (instead of a random sample) for a digital surrogate at the British Library.

Although addressing this question is beyond the scope of this article, we did observe, during the course of our investigation, a conspicuous pattern in the data which may help us begin to understand what is happening. We describe this pattern briefly in this section.

As a byproduct of our data collection process, our data include, for each title, whether or not the title has a digital surrogate from the British Library. These annotations permit us to ask whether or not there is evidence of digitization practices at the British Library which could be contributing to the differential availability across novel categories. (Because the British Library is a legal deposit library, we assume that collecting practices do not influence digital surrogate availability since the British Library received a copy of all novels.) Table 2 summaries the availability for the titles.

To assemble the data for Table 2 we did some additional data collection focused on the British Library. We looked for digital surrogates for an additional 30 randomly sampled 1836 novels. We only looked for digital surrogates at  the British Library. We did not check to see if these novels had surrogates at other digital libraries. Table 2 reports on digital surrogate availability at the British Library for these 1836 novels and all 1838 novels.

We find, much to our surprise, considerable evidence of a digitization practice at the British Library which discriminates on the basis of format: the British Library did not scan *any* multivolume novels from 1836. Although we were aware that very large (and very small) books were skipped by many digitization projects due to equipment being designed for books of certain (standard) sizes, we did not anticipate that multivolume works would be omitted from bulk digitization. The absence of digital surrogates for 1836 multivolume novels is unlikely to be due to chance. If the rate at which 1836 novels at the British Library have digital surrogates is 22.6% —the overall rate for 1836 titles, ignoring format—then the probability that 0 of the 37 multivolume novels would have been digitized is 1 in 13,000. We have no explanation for this pattern.

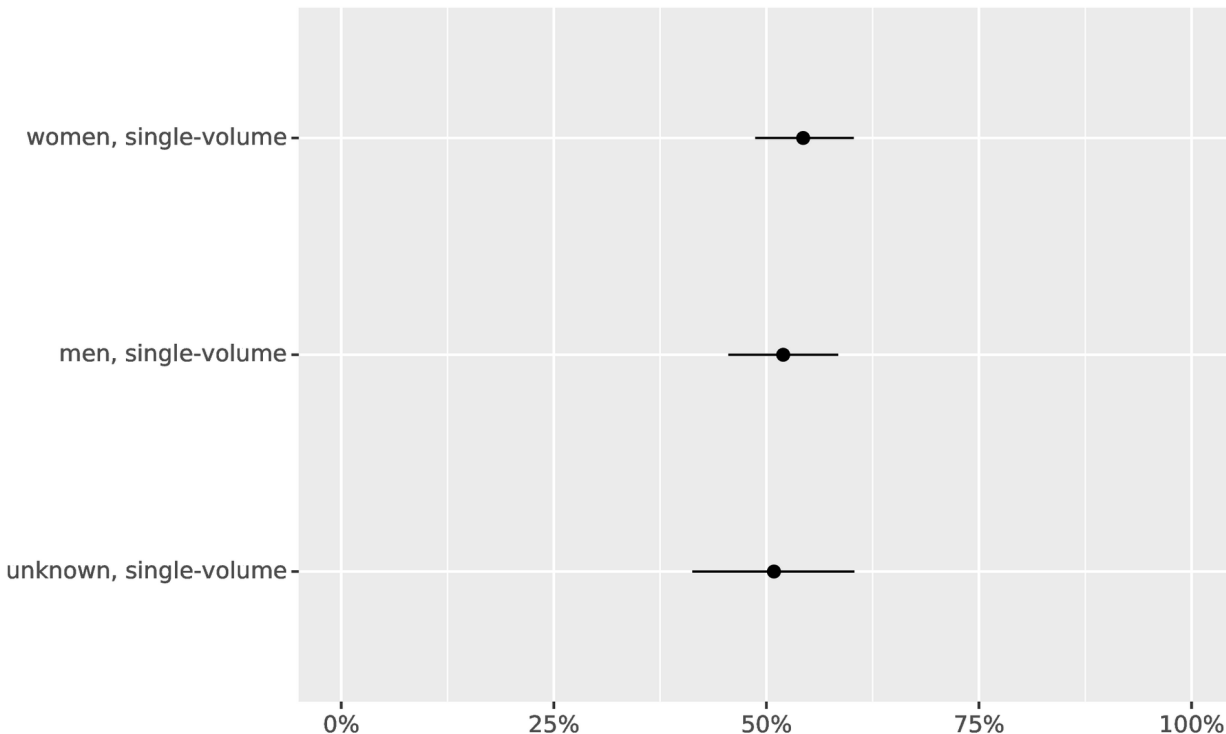

**Figure 2:** Estimated rates of British Library digital surrogate availability for single-volume novels published in 1836 and 1838. Points indicate posterior means, bars show 50% credible intervals.

Author gender, by contrast, does not predict digital surrogate availability in the British Library. Restricting our attention to single-volume novels, we find that digital surrogate availability does not appear to vary by gender in the British Library (Figure 2). (We model availability rates using the model described in Section 3.1, reducing the number of categories from six to three.) This result is consistent with the belief that (single-volume) novels in the British Library were indeed digitized at random. Because digitization procedures at different libraries resemble each other—in some cases, the firm or organization contracted to do the digitization is the same—this finding may offer weak support for the belief that digitization practices at other libraries did not discriminate among books in their holdings on the basis of author gender.

## 6 Discussion

This paper demonstrates the feasibility of studying the coverage of digital libraries using an exhaustive list of published documents (here, novels). That this approach improves on the strategy of evaluating the coverage of a digital library by reference to another larger collection or catalog of unknown comprehensiveness is easy to appreciate. An exhaustive list of published books is a stable reference point. Given such a list, comparing different collections is straightforward: each collection has some calculable percentage of the documents in the exhaustive list. Determining the share of documents not present in any of the collections is also straightforward. Comparing several collections without such a fixed point is, by contrast, complicated. A given digital library may have impressive coverage with respect to one collection but not another. Although an ersatz reference point in such a case would be the union of the items in the several collections, such a reference point is not usable in future work which attempts to compare new collections which contain previously unseen items. Using an exhaustive list avoids this complication. Future research on the coverage of digital libraries should seek, whenever possible, to use fixed reference points such as exhaustive lists.

Our work also suggests a broader research program, the data-intensive study of trans-Atlantic library collection practices. Digital surrogate availability permits distinguishing between export editions and first editions which have been transported across the Atlantic. Export editions aside, very few copies of first editions of books printed in the British Isles and intended for audiences in the British Isles were transported across the Atlantic Ocean. The overwhelming majority remained in the British Isles. The existence of one such book in a North American library—revealed by the existence of a digital surrogate—is a signal that someone—perhaps a library donor or individual responsible for expanding a collection—valued the book enough to finance its transportation across more than 2500 nautical miles. The existence of several such books is an even stronger signal. For researchers interested in the reception of a book, this signal offers evidence of non-negligible readership, analogous to the way the existence of a second edition hints that (a publisher believed) a book sold well in its initial outing. As every digital surrogate available from the major digital libraries indicates the contributing library, the necessary information for this genre of research is readily available.

A more immediate need, however, is a fuller account of which novels are more likely to have digital surrogates. Are 19th-century novels by women and novels by authors of unknown gender always less likely to have digital surrogates? Or is the phenomenon confined to the first half of the 19th century? Does novel format strengthen or attenuate this tendency? Answering these questions will produce a more complete narrative of how the major digital libraries' holdings differ from the population of 19th-century novels. This, in turn, will yield clues about which library collecting and digitization practices contribute to the differential availability of digital surrogates.

## 6.1 Limitations

The primary limitation of our study is that it provides information about digital libraries' coverage of new prose fiction published during the late 1830s. The digital libraries' holdings of non-fiction books and books (fiction and non-fiction) published during other periods may indeed reflect the population. Moreover, as our study narrowly concerns previously unpublished prose fiction, it may be the case that digital libraries' holdings of subsequent editions (i.e., editions other than first editions) published in the late 1830s may reflect the population of published subsequent editions. There are, to our knowledge, no systematic studies of collecting and digitization practices that permit us to generalize our findings beyond the late 1830s.

Another weakness of the study is that it does not distinguish between physical holdings and digital surrogates. A more nuanced investigation would have identified, for each library contributing digital surrogates, which books were held but not digitized. Such an investigation would provide insight into the collection and digitization practices of specific libraries. Our study avoided doing this because we assumed—based on experience working on bibliographic projects—that the legal deposit libraries of Oxford and the British Library possessed copies of every novel. This assumption may merit revisiting. Some novels which were once in the collections of the legal deposit libraries may have been lost or destroyed.

## 7 Conclusion

In this paper we describe the availability of digital surrogates of novels first published in the British Isles during the late 1830s. We compare the set of novels which have at least one digital surrogate available from the major digital libraries with the population of novels published during the period. We find that digital surrogate availability differs by author gender and format. Multivolume novels by men are most likely to have at least one digital surrogate. Single-volume novels, novels by women, and novels by authors of unknown gender are less likely to have digital surrogates. Future research may offer an account of the causes of the differential availability of surrogates. We speculate that library collecting practices play a role—for instance, a library may tend to collect novels in subgenres in which men authors predominate. Equally so, library digitization practices may play a role in the availability of surrogates—for instance, the British Library appears to have excluded multi-volume novels published in 1836 from bulk digitization efforts.

The period under examination, the late 1830s, was chosen because it includes the most recent years for which exhaustive bibliographies exist. We have no reason to believe the late 1830s was exceptional in the relevant sense. A novel being published in the late 1830s seems unlikely, by itself, to influence the likelihood that the novel will have been collected or digitized by a library. Moreover, the existence of conspicuous biases in collecting or digitization practices for novels makes it more likely that biases also exist which influence the collecting of non-fiction works. Absent evidence to the contrary, it seems prudent for researchers to assume that gathering samples of fiction or non-fiction works published in the British Isles during the 1830s and 1840s from the major digital libraries will yield samples which do not reflect the population.

## References

About Google Books. (2011). http://books.google.com/googlebooks/history.html.

Alexander, I. (2010, March 3). Copyright Law and the Public Interest in the Nineteenth Century. Bloomsbury Publishing. eprint: KwncBAAAQBAJ

Bamman, D., Carney, M., Gillick, J., Hennesy, C., & Sridhar, V. (2017). Estimating the date of first publication in a large-scale digital library. In Proceedings of the 17th ACM/IEEE Joint Conference on Digital Libraries (pp. 149–158). IEEE Press.

Bassett, T. (n.d.). At the Circulating Library. https://www.victorianresearch.org/atcl/.

Bassett, T. J. (2008). The Production of Three-Volume Novels, 1863-1897. Papers of the Bibliographical Society of America, 102(1), 61–75. Retrieved August 18, 2011, from http://users.ipfw.edu/bassettt/research.html

Bode, K. (2017, March 1). The Equivalence of "Close" And "Distant" Reading; Or, toward a New Object for Data-Rich Literary History. Modern Language Quarterly, 78(1), 77–106. doi:10.1215/00267929-3699787

Boyd, A. E. (1998). "What! Has she got into the 'Atlantic?'" Women Writers, the Atlantic Monthly, and the Forming of the American Canon. American Studies, 39(3), 5–36. doi:10.1353/amsj.v39i3.2695

Carpenter, B., Gelman, A., Hoffman, M., Lee, D., Goodrich, B., Betancourt, M., … Riddell, A. (2017). Stan: A Probabilistic Programming Language. Journal of Statistical Software, 76(1), 1–32. doi:10.18637/jss.v076.i01

Feather, J. (1994). Publishing, Piracy and Politics: An Historical Study of Copyright in Britain. New York, NY: Mansell.

Garside, P., Berlanger, J., & Mandal, A. (2001). The English Novel, 1800–1829: Update 1 (Apr 2000–May 2001). Retrieved September 28, 2015, from http://www.british-fiction.cf.ac.uk/guide/update1.html

Garside, P., Mandal, A., Ebbes, V., Koch, A., & Schöwerling, R. (2004). The English Novel, 1830–1836: A Bibliographical Survey of Fiction Published in the British Isles. http://www.cardiff.ac.uk/encap/journals/corvey/1830s/1830-36.html.

Garside, P., & Schöwerling, R. (2000). The English Novel, 1770-1829: A Bibliographical Survey of Prose Fiction Published in the British Isles (P. Garside, J. Raven, & R. Schöwerling, Eds.). Oxford: Oxford University Press.

HathiTrust Digital Library. (n.d.). Member Community. Retrieved February 11, 2020, from https://www.hathitrust.org/community

Hoffelder, N. (2013, July 9). Internet Archive Now Hosts 4.4 Million eBooks, Sees 15 Million eBooks Downloaded Each Month - The Digital Reader. Retrieved October 31, 2019, from https://web.archive.org/web/20131110091506/http://www.the-digital-reader.com/2013/07/09/internet-archive-now-hosts-4-4-million-ebooks-sees-15-million-ebooks-downloaded-each-month/

Jones, E. (2011). Google Books as a general research collection. Library Resources & Technical Services, 54(2), 77–89.

Kahle, B. (2008, May 26). Internet Archive Forums: Books Scanning to be Publicly Funded. Retrieved October 31, 2019, from https://archive.org/post/194217/books-scanning-to-be-publicly-funded

Kahle, B. (2014, January 14). Public Access to the Public Domain: Copyright Week. Retrieved February 11, 2020, from https://blog.archive.org/2014/01/14/public-access-to-the-public-domain-copyright-week/

Lauterbach, C. E., & Lauterbach, E. S. (1957, December). The Nineteenth Century Three-Volume Novel. The Papers of the Bibliographical Society of America, 51(4), 263–302. doi:10.1086/pbsa.51.4.24299448

Open Content Alliance (OCA) - ContributorS. (2008, February 5). Retrieved October 31, 2019, from https://web.archive.org/web/20080205024016/http://www.opencontentalliance.org/contributors.html

Raven, J., & Forster, A. (2000). The English novel, 1770-1829: A Bibliographical Survey of Prose Fiction Published in the British Isles (P. Garside, J. Raven, & R. Schöwerling, Eds.).

Oxford: Oxford University Press.

Riddell, A., Bassett, T. J., Schneider, L., Mills, H., Yarnell, A., Condon, R., … Duke, S. (2019, September 5). Common Library 1.0: A Corpus of Victorian Novels Reflecting the Population in Terms of Publication Year and Author Gender. arXiv: 1909.02602 [cs]. Retrieved October 18, 2019, from http://arxiv.org/abs/1909.02602

Sare, L. (2012). A comparison of HathiTrust and Google Books using federal publications. Practical Academic Librarianship: The International Journal of the SLA Academic Division, 2(1), 1–25.

Tuchman, G. (1989). Edging Women Out: Victorian Novelists, Publishers, and Social Change. New Haven: Yale University Press.

Wu, T. (2015). What Ever Happened to Google Books? *New Yorker*, September 12, 2015. https://www.newyorker.com/business/currency/what-ever-happened-to-google-books.

## Appendix A: Gender Annotations

The gender annotation procedure used in both exhaustive bibliographies is believed to be essentially the same. There is one minor difference which only concerns novels which are written by an unknown author.

If the historical individual who wrote the novel is *known*, their gender is used. Thanks to the labors of literary historians, bibliographers, and genealogists, the historical individual is typically known. If, however, the historical individual who wrote the novel is *unknown*, a gender annotation different than "unknown" will be made if one of the following conditions are met:

- The author's name (or pseudonym) is strongly associated with individuals of one gender rather than another. For example, a novel by a "Lady of Rank" would be coded as a woman-authored novel. Example: *The Glanville Family* (1838).

- The author indicates their gender in paratextual material such as a preface. For example, a non-narrative preface by the author may indicate the author's gender by the use of gendered pronouns.

This method of arriving at author gender annotations strongly resembles the method used by Garside and Schöwerling (2000). In handling novels whose authors are unknown, we depart from that method in one respect: whereas Garside and collaborates tend to only use information appearing on the title page, we also use information appearing in non-narrative prefaces.

[1] *About Google Books*, http://books.google.com/googlebooks/history.html, June 2011.

[2] Katherine Bode, "The Equivalence of "Close" And "Distant" Reading; Or, toward a New Object for Data-Rich Literary History," *Modern Language Quarterly* 78, no. 1 (March 1, 2017): 77–106, doi:10.1215/00267929-3699787.

[3] Isabella Alexander, *Copyright Law and the Public Interest in the Nineteenth Century* (Bloomsbury Publishing, March 3, 2010), 62-63.

[4] *About Google Books.*

[5] Tim Wu, "What Ever Happened to Google Books?" [In en], September 2015, issn: 0028-792X.

[6] "Open Content Alliance (OCA) - ContributorS," February 5, 2008, https://web.archive.org/web/20080205024016/http://www.opencontentalliance.org/contributors.html.

[7] Brewster Kahle, "Internet Archive Forums: Books Scanning to Be Publicly Funded," May 26, 2008, https://archive.org/post/194217/books-scanning-to-be-publicly-funded.

[8] Nate Hoffelder, "Internet Archive Now Hosts 4.4 Million eBooks, Sees 15 Million eBooks Downloaded Each Month - The Digital Reader," July 9, 2013, https://web.archive.org/web/20131110091506/ http://www.the-digital-reader.com/2013/07/09/internet-archive-now-hosts-4-4-million-ebooks-sees-15-million-ebooks-downloaded-each-month/.

[9] HathiTrust Digital Library, "Member Community," https://www.hathitrust.org/community.

[10] For texts written in languages other than English the holdings of these digital libraries are marginal by comparison to other digital libraries (e.g., Norway's Nasjonalbiblioteket, France's Bibliothèque Nationale de France, and Germany's Bayerische Staatsbibliothek).

[11] Laura Sare, "A Comparison of HathiTrust and Google Books Using Federal Publications," *Practical Academic Librarianship: The International Journal of the SLA Academic Division* 2, no. 1 (2012): 1–25.

[12] Brewster Kahle, "Public Access to the Public Domain: Copyright Week," January 14, 2014, https://blog.archive.org/2014/01/14/public-access-to-the-public-domain-copyright-week/.

[13] David Bamman et al., "Estimating the Date of First Publication in a Large-Scale Digital Library," in *Proceedings of the 17th ACM/IEEE Joint Conference on Digital Libraries* (IEEE Press, 2017), 149–158.

[14] Edgar Jones, "Google Books as a General Research Collection," *Library Resources & Technical Services* 54, no. 2 (2011): 77–89.

[15] Ibid.

[16] Sare, "A Comparison of HathiTrust and Google Books Using Federal Publications."

[17] Ibid.

[18] John Feather, *Publishing, Piracy and Politics: An Historical Study of Copyright in Britain* (New York, NY: Mansell, 1994), 97, isbn: 978-0-7201-2135-3.

[19] Personal communication with Daniel Wilson (The Alan Turing Institute) on November 16, 2019. The Alan Turing Institute is located in and affiliated with the British Library.

[20] Although information about the print run of specific editions is typically lost, we believe there may be serviceable proxies available. For example, a book which a publisher anticipated being popular may have been more likely to be printed using the "prestige" three-volume ("triple decker") format.

[21] Charles E. Lauterbach and Edward S. Lauterbach, "The Nineteenth Century Three-Volume Novel," *The Papers of the Bibliographical Society of America* 51, no. 4 (December 1957): 269, issn: 0006-128X, 2377-6528, doi:10.1086/pbsa.51.4.24299448, https://www.journals.uchicago.edu/doi/10.1086/pbsa.51.4. 24299448; Troy J. Bassett, "The Production of Three-Volume Novels, 1863-1897," *Papers of the Bibliographical Society of America* 102, no. 1 (2008): 74-75, http://users.ipfw.edu/bassettt/research.html.

[22] Gaye. Tuchman, *Edging Women Out: Victorian Novelists, Publishers, and Social Change* (New Haven: Yale University Press, 1989), isbn: 0300043163 (alk. paper); Anne E. Boyd, ""What! Has She Got into the 'Atlantic?'" Women Writers, the Atlantic Monthly, and the Forming of the American Canon," *American Studies* 39, no. 3 (September 1998): 5–36, issn: 0026-3079, doi:10.1353/amsj.v39i3.2695.

[23] Troy J. Bassett, *At the Circulating Library*, https://www.victorianresearch.org/atcl/.

[24] James Raven and Antonia Forster, *The English Novel, 1770-1829: A Bibliographical Survey of Prose Fiction Published in the British Isles*, ed. Peter Garside, James Raven, and Rainer Schöwerling, vol. 1 (Oxford: Oxford University Press, 2000), isbn: 978-0-19-818318-1; Peter Garside and Rainer Schöwerling, *The English Novel, 1770-1829: A Bibliographical Survey of Prose Fiction Published in the British Isles*, ed. Peter Garside, James Raven, and Rainer Schöwerling, vol. 2 (Oxford: Oxford University Press, 2000), isbn: 978-0-19-818318-1.

[25] Verifying that a digital surrogate exists takes a considerable amount of time. The catalogs of each digital library must be searched separately. Even when a surrogate is found which appears to be a match for the given novel (i.e., the title, author, and publication year match), page images must be inspected. It is common to find a North American or French edition of a novel which is published in the same year as the first edition. Verifying that all volumes of a multivolume work are present is also time consuming. Oxford, for example, frequently binds together the separate volumes of a multivolume novel into one "volume". So a single "volume" of a multivolume novel at Oxford needs to be carefully checked to verify it is complete. Our use of a random sample of 1836 novels reflects a desire to economize on time.

[26] Bassett, *At the Circulating Library*.

[27] The average number of novels published in 1836 and 1838 is 92. If we assume that 92 novels were published in 1835, 1837, and 1839, then the size of the late-1830s population is 460 novels.

[28] These novels are included in a series of six updates to the bibliography (e.g., (Garside, Berlanger, & Mandal, 2001). The number of novels "discovered" is very small relative to the size of the bibliography. (The 1800-1829 bibliography has more than 2,000 titles.) For example, the first update features 10 newly discovered novels.

[29] Bob Carpenter et al., "Stan: A Probabilistic Programming Language," *Journal of Statistical Software* 76, no. 1 (2017): 1–32, issn: 1548-7660, doi:10.18637/jss.v076.i01.